\documentclass[conference]{IEEEtran}
\IEEEoverridecommandlockouts
\usepackage{cite}
\usepackage{graphicx}
\usepackage{xcolor}
\usepackage{amsmath,amssymb,amsfonts}
\usepackage{amsthm}
\usepackage{dsfont}
\usepackage{enumitem}

\newcommand{\todo}[1]{{\leavevmode\color{black}#1}}

\usepackage{url}

\usepackage[pscoord]{eso-pic}
\newcommand{\placetextbox}[3]{
  \setbox0=\hbox{#3}
  \AddToShipoutPictureFG*{
    \put(\LenToUnit{#1\paperwidth},\LenToUnit{#2\paperheight}){\vtop{{\null}\makebox[0pt][c]{#3}}}%
  }%
}%

\def\BibTeX{{\rm B\kern-.05em{\sc i\kern-.025em b}\kern-.08em
    T\kern-.1667em\lower.7ex\hbox{E}\kern-.125emX}}

\begin{document}
\placetextbox{0.5}{0.98}{\texttt{This work has been submitted to the IEEE for possible publication. Copyright may be }}%
\placetextbox{0.5}{0.964}{\texttt{transferred without notice, after which this version may no longer be accessible.}}%

\title{Untangling Dense Crowds with mmWave Radar: \\
From Crowd Semantics to Individual Spatial Behaviors}
\author{Aaditya Prakash Kattekola, Anurag Pallaprolu, and Yasamin Mostofi
\thanks{Aaditya Prakash Kattekola, Anurag Pallaprolu and Yasamin Mostofi are with the Department of Electrical and Computer Engineering, University of California Santa Barbara, CA, 93106 USA (e-mail: aadityaprakash@ucsb.edu, apallaprolu@ucsb.edu, ymostofi@ece.ucsb.edu).This work is funded in part by ONR award N00014-23-1-2715 and in part by NASA award 80NSSC25M7102.}}

\maketitle

\begin{abstract}
In this paper, we present a novel framework for recovering individual footprints and pedestrian spatial behaviors in dense crowds. This is a highly challenging problem, as pedestrians can merge or block one another often and for prolonged periods in crowded areas, leading to strongly entangled observations. Our approach jointly leverages learned crowd semantics and principled reasoning over persistent radar observability loss to systematically infer individual spatial footprints. At the macroscopic level, we learn spatial usage patterns of the crowd from radar point clouds, while at the microscopic level we develop a physics-informed model of radar observability loss caused by sustained merging and occlusion. Building on this macro-micro structure, we introduce a semantic-guided multi-hypothesis reasoning foundation that evaluates competing hypotheses via principled reasoning over many-to-one and one-to-none observation mappings, enabling untangling of individual spatial footprints. \todo{We evaluate the framework through 18 real-world experiments using an off-the-shelf mmWave radar, involving crowds of up to (and including) 10 individuals across six categories of diverse pedestrian behaviors, and four environments. Our methodology robustly recovers individual spatial footprints, even during prolonged and persistent observability loss, demonstrating strong structural consistency with the ground-truth. The recovered footprints further enable accurate zonal analysis of pedestrian space usage. Finally, we present an ablation study and further compare against state-of-the-art. The proposed foundation substantially outperforms the state-of-the-art across all experiments, achieving an average median DTW of 0.31 m compared to 0.96 m, highlighting the framework's robustness to frequent and prolonged merging and occlusion.}

\end{abstract}



\section{Introduction}
\label{sec:introduction}

\IEEEPARstart{U}NDERSTANDING how individuals use a space is important for several emerging applications. Recovering individual footprints, for instance, can unlock actionable insights, including detecting unusual or unsafe  behaviors in public spaces~\cite{kratz2009anomaly},  personalized assistance in complex environments such as airports or hospitals, and characterizing how individuals respond to spatial constraints and environmental structure. It can further support congestion analysis~\cite{zanlungo2023pure, feliciani2018measurement} or reveal how retail layouts shape pedestrian flow~\cite{li2022retail, torrens2022data}, addressing problems of practical consequence.

\begin{figure*}
    \centering
    \includegraphics[width=\linewidth]{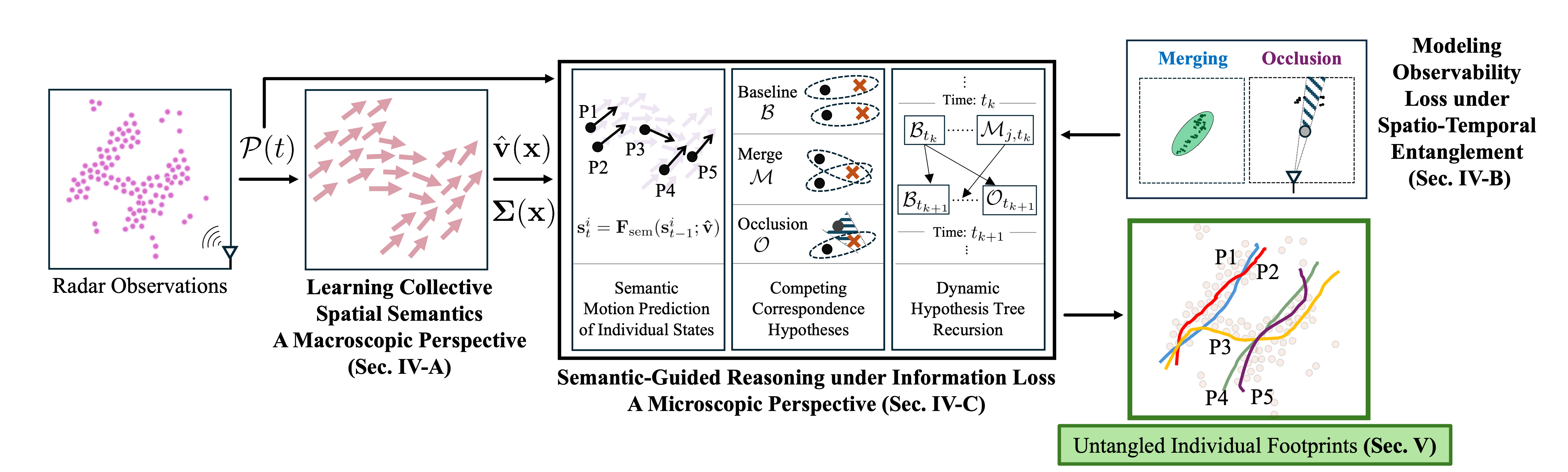}
     \vspace{-21pt}
\caption{High-level overview of the proposed methodology for untangling individual footprints in dense crowds. The framework first learns collective spatial semantics (IV-A) and develops a principled model of radar observability loss due to merging and occlusion (IV-B), then systematically generates and evaluates competing hypotheses to resolve observation ambiguity and recover individual spatial footprints and behaviors (IV-C).}
  \vspace{-20pt}
    \label{fig:pipeline_flowchart}
\end{figure*}

Recovering individual spatial footprints and patterns of space usage, however, is inherently challenging due to the complex and interaction-rich dynamics of human movement in crowded environments. In many urban settings, such as transit stations, shopping centers, and stadium concourses, large numbers of pedestrians continuously move through shared spaces. In these environments, individuals rarely move in isolation. Instead, they often walk in pairs or small groups whose movements can remain persistently coupled~\cite{moussaid2010walking, aveni1977not}. Their paths are further shaped by the physical layout of the environment~\cite{daamen2003experimental}, as pedestrians continuously adapt their motion to avoid collisions with obstacles and nearby individuals~\cite{fruin1987pedestrian, helbing1995social}. As a result, pedestrian motion is highly dynamic: individuals repeatedly approach, couple, decouple, 
and at times block one another as they navigate shared spaces, producing complex spatial behaviors that are difficult to disentangle at the level of each person.

Prior work has explored recovering individual spatial behaviors in crowded environments using several sensing modalities. Vision-based systems, for instance, are prominent owing to the rich information content of image and video streams, enabling individual separation through appearance cues~\cite{zhang2022bytetrack, yang2024hybrid}. However, they are fundamentally constrained by viewpoint limitations~\cite{li2025review}, sensitivity to lighting, and significant privacy concerns around public deployment~\cite{kansal2025implications, qandeel2024facial}. LiDAR inherits similar vulnerabilities as vision while introducing substantially higher hardware and deployment costs~\cite{torres2023pedestrian}. On the other hand, infrared-based systems offer a lighting-invariant alternative but operate at spatial resolutions that restrict them to coarse occupancy estimation rather than per-person footprint recovery~\cite{raykov2016predicting, shokrollahi2024passive}. Finally, acoustic approaches have explored footstep and motion-induced sound signatures, but remain limited to controlled environments with low ambient noise~\cite{cai2021we, wu2024afpild}. Over the past decade, WiFi signals have been widely leveraged for sensing tasks beyond communication~\cite{wei2025survey}, including imaging~\cite{yao2024wiprofile, shang2024liquimager, pallaprolu2022wiffract}, person identification~\cite{chen2025environment, wang2025muid, korany2020multiple}, and activity recognition~\cite{xu2025evaluating, miao2025wi, cai2020teaching}. While WiFi-based solutions have been explored for crowd analytics, the inherently low spatial resolution of commodity WiFi hardware confines such approaches to occupancy estimation~\cite{depatla2019occupancy}, or motion inference of simple scenarios~\cite{qian2018widar2}, insufficient for the individual-level footprint recovery that dense crowd settings demand. 

In recent years, mmWave has emerged as a compelling sensing alternative, driven by the rapid development of Integrated Sensing and Communication (ISAC) as a key component of future 6G networks~\cite{liu2022survey, liu2022integrated} and the growing availability of compact, low-cost commodity FMCW transceivers~\cite{awr2243boostug}. Prior work on mmWave-based crowd sensing largely 
falls into two categories. The first focuses on estimating aggregate descriptors of crowd behavior without resolving individual pedestrian behavior~\cite{pallaprolu2025mmflux, Zhang2024Sensors, vales2024iot, pallaprolu2024crowd, li2023indoor}. The second category, on the other hand, aims to recover individual footprints, but existing systems are mainly limited to small number of occupants (five or less) and spatially-sparse scenarios where pedestrian interactions are minimal and their paths remain largely well separated, typically following simple and prescribed routes~\cite{cheng2025space, chen2024mmtai, shamsfakhr2024multi, wang2025grouped, liu2024pmtrack}. More importantly, the existing work does not consider challenging cases where sustained merging or occlusion results in prolonged observability loss. Sec.~\ref{sec:related_work} provides a comprehensive treatment of prior work. 

As a result, recovering individual spatial footprints in dense, dynamic, and interaction-rich crowds where pedestrians often merge, separate, and block one another, leading to \textbf{prolonged periods of highly entangled observations}, remains an open challenge. \todo{We next summarize the key characteristics of natural crowds that make this problem particularly challenging.}

\subsection*{\todo{Key Challenges in Recovering Crowd Footprints}}
\todo{Natural crowded environments exhibit several characteristics that make recovering individual spatial footprints particularly challenging. We next highlight \textbf{key properties that are hallmarks of such environments} and must be addressed to successfully recover individual spatial footprints.} 

\textbf{Crowd Size and Local Density:} Many individuals traverse the sensing field of view (FOV) simultaneously, often moving in close proximity. This leads to locally dense clusters of pedestrians whose composition evolves over time.

\textbf{Observation Ambiguity from Merging and Occlusion:} As individuals move in close proximity, observations from different pedestrians frequently merge or become occluded. Radar returns from multiple individuals may thus overlap, or disappear due to blockage. As a result, a single observation may correspond to multiple individuals, while at other times an individual may produce no observable return.

\todo{\textbf{Prolonged Observability Loss:} In realistic scenarios, merging and occlusion among individuals often persist over extended time intervals rather than occurring sporadically, giving rise to prolonged spatio-temporal entanglement and extended periods of observability loss.}

\textbf{Interaction-Rich Motion:} Pedestrian motion is inherently interdependent, as individuals continuously adjust their movement in response to other pedestrians and obstacles. These interactions produce intermittent grouping, avoidance maneuvers, and prolonged periods during which individual footprints remain closely coupled.

\textbf{Natural Space Usage:} Pedestrians exhibit natural patterns of space usage, rather than following scripted routes.

These properties give rise to what we term spatio-temporal entanglement in this paper.

\textbf{\textit{Spatio-Temporal Entanglement}:} We define \textit{Spatio-Temporal Entanglement} as the phenomenon in which radar point cloud observations associated with distinct individuals become intertwined across space and time, often for prolonged durations, making it difficult to disentangle and recover each individual’s footprint from mmWave point clouds.

\todo{In this paper, we introduce a framework that recovers the spatial footprints and space usage patterns of individuals in dense, interaction-rich crowds using off-the-shelf mmWave radar. Our approach methodically reasons over ambiguous radar observations to disentangle individual behaviors despite frequent and prolonged merging and/or occlusion which can induce long-lasting observability entanglement and loss. Fig.~\ref{fig:pipeline_flowchart} provides an overview of our pipeline.}

We next summarize our key contributions.

\textbf{Statement of Contributions:}

{\noindent\textbf{a)} We introduce a novel framework that enables the recovery of individual spatial behaviors from spatio-temporally entangled mmWave radar point clouds in dense crowds, even when such entanglements persist over long periods of time. By jointly leveraging learned crowd semantics, observability loss modeling, and principled reasoning over radar observability loss, our framework systematically untangles overlapping radar observations to recover individual spatial footprints and behaviors.

\noindent\textbf{\todo{b)}} \todo{The proposed framework adopts a macro-to-micro modeling paradigm that bridges crowd semantics and radar observability modeling. At the macroscopic level, it learns spatial usage patterns of the crowd from radar point clouds, while at the microscopic level it develops a physics-informed model of radar observability loss caused by merging and occlusion, guiding reasoning over individual states.}

\noindent\textbf{\todo{c)}} \todo{Building on this macro-to-micro modeling structure, we propose a principled semantic-guided multi-hypothesis reasoning framework. The method explicitly models many-to-one (merging) and one-to-none (occlusion) mappings between predicted states and observations. It then analytically generates, scores, and propagates the corresponding hypotheses using semantic-guided state predictions and principled radar observability modeling.}

\noindent\textbf{d)} 
We propose novel metrics to quantify the degree of spatio-temporal crowd entanglement. These metrics provide a principled measure of observation ambiguity in dense crowds and systematically characterize the inherent difficulty of disentangling individual behaviors, a challenge that has largely lacked quantitative treatment.
 
\noindent\textbf{e)} We extensively evaluate the proposed framework through 18 real-world experiments conducted using a TI AWR2243BO\-OST off-the-shelf mmWave radar~\cite{awr2243boostug}. Our experiments involve crowds of up to (and including) 10 individuals spanning six categories of pedestrian space usage behaviors, and 4 different environments including both indoor and outdoors.}

Our framework robustly recovers individual space usage patterns, with an average median DTW of 0.31 m across all experiments, demonstrating significant improvement over the state-of-the-art. The recovered footprints exhibit strong structural agreement with camera-derived ground truth, validating the framework’s effectiveness in untangling individual behaviors in challenging, interaction-rich crowds, that can experience several instances of prolonged observability loss. We further evaluate the accuracy of inferred space usage through zonal occupancy analysis, providing quantitative insights into individual spatial usage patterns.

\section{Related Work}
\label{sec:related_work}
RF sensing has emerged as a promising modality for crowd analytics, as it does not depend on lighting conditions, alleviates privacy concerns, and avoids infrastructural overhead associated with alternative modalities discussed in Sec. \ref{sec:introduction}. Prior RF-based approaches to crowd sensing span a hierarchy of inference granularity, from coarse statistics to fine-grained individual footprint recovery. Early work has primarily relied on active sensing in the sub-6GHz bands, using WiFi probe requests or BLE beacons, to estimate crowd occupancy and density~\cite{schauer2014estimating, vattapparamban2016indoor, longo2019accurate, stanciu2023privacy}. These approaches require individuals to carry actively transmitting devices and typically provide only aggregate statistics about the crowd. While passive RF sensing addresses this limitation, the limited spatial resolution of sub-6GHz hardware restricts reliable inference of individual-scale footprints to scenarios involving only a small number of spatially well-separated targets~\cite{zhu2024commodity, wu2021witraj, karanam2019tracking, qian2018widar2}.

mmWave sensing overcomes these limitations through higher spatial resolution, enabled by the larger available bandwidth. Prior work in mmWave-based crowd sensing broadly falls into two categories. 

The first category focuses on aggregate-level crowd characterization. For instance, prior work has leveraged commodity mmWave radars for occupancy estimation~\cite{vales2024iot, ren2023grouped, li2023indoor, pallaprolu2024crowd, instruments2020people}, as well as for characterizing spatial flow fields~\cite{Ding2024milliFlow, Ding2022RaFlow, zhao2025performance, zhai2025dmrflow}. More recently, flow field representations have been extracted directly from mmWave point clouds to characterize dominant motion directions and flow separators in structured crowd settings~\cite{pallaprolu2025mmflux, Zhang2024Sensors, pallaprolu2026mmridge}. These approaches, however, are not suited for recovering individual spatial footprints, focusing instead on aggregate crowd metrics.

The second category seeks to recover individual motion footprints from mmWave point clouds through the lens of multi-target tracking. However, these approaches are typically developed for relatively sparse environments and are not suited for dense, interaction-rich crowd settings~\cite{liu2024pmtrack, cheng2025space, chen2024mmtai, chen2023environment, wang2025grouped, shamsfakhr2024multi}. More specifically, prior work in the second category share the following assumptions that break down in spatio-temporally entangled crowds settings:

\noindent\textbf{1)} They predominantly consider small groups of five or fewer individuals traversing well-separated routes with minimal mutual interaction. Under such conditions, individual motion is approximately linear, radar observations remain spatio-temporally separable, and occlusion events are 
transient. 

\noindent\textbf{2)} Radar observations are assumed to map one-to-one with individual motion predictions~\cite{shamsfakhr2024multi, sathyan2013multiple, zhu2024joint, dai2023interpersonal}. This assumption severely breaks down if individuals move in close proximity, causing radar returns from multiple people to merge into a single observation (many-to-one case). Moreover, blockage cases can result in one-to-none observations. More importantly, existing work assumes that such cases are short-lived and transient. For instance, line-of-sight (LOS) occlusion is treated as a brief, intermittent outage bridged by motion prediction in the absence of a corresponding radar observation~\cite{segal2018occlusion, kong2022m3track}. In realistic dense crowds, however, merging and occlusion can persist for extended periods, giving rise to prolonged spatio-temporal entanglement and observability loss. Without explicitly modeling these phenomena, existing approaches can incur systematic ambiguities in associating radar observations with individuals.

\noindent\textbf{3)} Individual motion is modeled through constant-velocity kinematics, with trajectories designed to be random walks across the FOV~\cite{shamsfakhr2024multi, liu2024pmtrack}. This abstraction cannot capture the inherently nonlinear, socially-coupled nature of real pedestrian motion, where individuals continuously adjust their paths in response to neighboring pedestrians rather than moving independently at random~\cite{dachner2022visual, helbing1991mathematical, moussaid2010walking}. 

To the best of our knowledge, no prior work has addressed these limitations to enable individual footprint recovery in dense, interaction-rich crowds, which is the main motivation for the proposed framework of this paper. 

\section{Radar Observations in Dense Crowds}
\label{sec:radar_obs_in_dense_crowds}
We next describe the front-end pipeline that generates point clouds, denoted by $\mathcal{P}(t)$, from raw radar data, and present metrics to characterize spatio-temporal entanglement in real point clouds, empirically motivating our proposed framework for untangling dense crowd observations in Sec.~\ref{sec:proposed_solution}.

\textbf{Point Cloud Generation:} We consider an FMCW MIMO radar with $N_{\text{TX}}$ transmit (TX) and $N_{\text{RX}}$ receive (RX) antennas operating in a TDM-MIMO configuration. Each TX emits a linear chirp with instantaneous frequency $f_{\text{TX}}(t)=f_0+St$, where $f_0$ is the carrier frequency, $B$ is the bandwidth, $T_c$ is the chirp duration, and $S=B/T_c$ is the chirp slope. Reflections off of individuals in the scene are mixed with the transmitted signal at each RX to produce intermediate-frequency (IF) signals~\cite{stove1992linear, rao2017introduction}, which are discretized over ADC samples and chirps. The IF signals from all TX-RX pairs are reorganized into a 2D virtual planar array, and we denote the discretized IF signal at virtual array element $(p,q)$ by $s_{\text{IF}}^{(p,q)}[m,n]$, where $m, n$ index chirps and ADC samples respectively. 

Point cloud extraction proceeds in two stages. First, an FFT is applied on the IF signals along the ADC dimension $n$ to obtain per-chirp range profiles. The phase spectrum of the resulting range-FFT is then evaluated over sliding windows of chirps to detect range bins exhibiting spectral bandwidth that is consistent with human motion~\cite{pallaprolu2024crowd}. This step effectively suppresses static clutter and avoids CFAR-style thresholding~\cite{rohling2011ordered} over entire range–angle spectra, yielding a set of candidate sensing depths over time. Second, azimuthal angle-of-arrival (AOA) estimation is performed only at the detected ranges. For each virtual array element, a range–Doppler transform is computed over the corresponding chirp window, followed by a 2D FFT across the virtual aperture~\cite{pallaprolu2025mmflux}. 
Doppler bins are averaged for AOA estimation, and the elevation dimension is collapsed due to the limited vertical resolution when compared to the resolution in the azimuthal dimension. The resulting azimuthal spectrum is used to construct a 2D point cloud of radar measurements by selecting the bearing angle that maximizes the angular response at each detected range, with radial velocity extracted from the peak Doppler bins. The aforementioned process is repeated for each time step to obtain a temporally evolving point cloud representation for the crowd, i.e., $\mathcal{P}(t)$. At each time step, radar point cloud $\mathcal{P}(t)$ is clustered using a density-based approach \cite{ester1996density}, a common step to suppress sparse clutter. The resulting cluster centroids then provide radar observations that serve as inputs for our proposed framework.

\textbf{Statistical Characterization of Spatio-Temporal Entanglement:} We next propose two statistical measures derived directly from radar observations to quantify the extent of spatio-temporal entanglement across our experiments. The Observability Factor is defined as the ratio of the number of observations detected at a given time step (i.e., not occluded or merged) to the total crowd size, averaged over time. This captures the aggregate observability loss arising from both merging and occlusion where multiple individuals may yield a single merged observation, or an occluded individual may produce none at all. The Merge Factor, on the other hand, is defined as the number of observations at time step $t$ that fall within a human radius~\cite{biacromialCDC} of at least one observation at time step $t+1$, normalized by the crowd size and averaged over time. This provides a more nuanced measure of the extent to which observations participate in merge events as the crowd evolves. We note that these two measures are not mutually exclusive: a merge event necessarily contributes to observability loss, and thus the Observability Factor captures both merging and occlusion effects, while the Merge Factor isolates the merging component alone. 

\begin{figure}
    \centering
    \includegraphics[width=\linewidth]{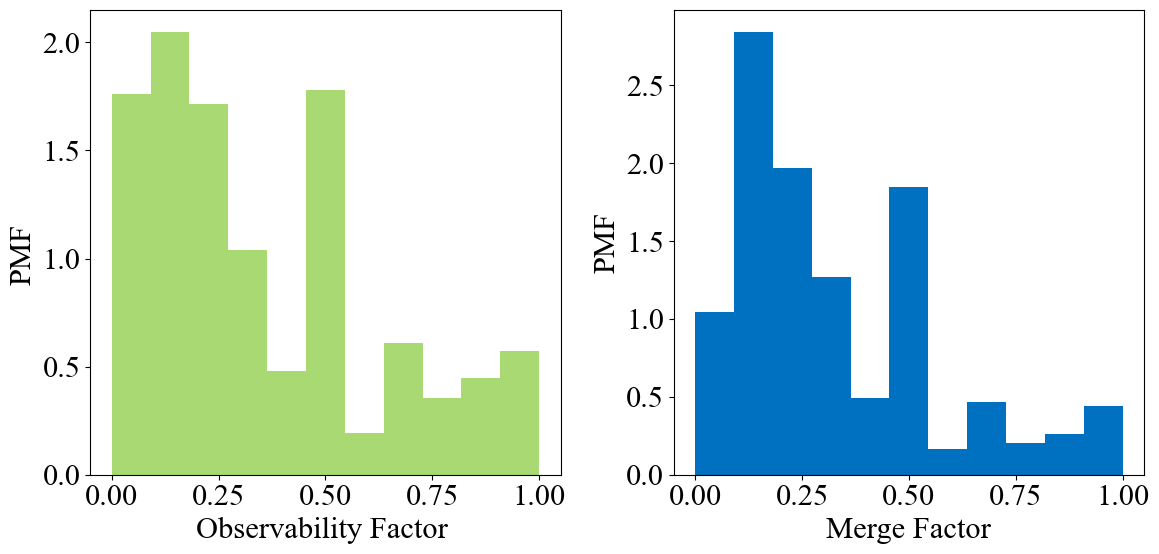}
    \vspace{-20pt}
    \caption{Proposed spatio-temporal entanglement metrics characterizing observation ambiguity over all our experiments. Histogram of (Left) Observability Factor (fraction of observations not occluded or merged), (Right) Merge Factor (fraction of observations involved in merge events).}
    \vspace{-18pt}
    \label{fig:STEI}
\end{figure}

The distributions of both factors across our dataset of dense crowd evolutions are shown in Fig.~\ref{fig:STEI}. More specifically, Fig.~\ref{fig:STEI} (Left) demonstrates that all our experiments exhibit significant loss of observability, with an average Observability Factor of 0.36 i.e., in a 10-person crowd, only 4 individuals produce distinct radar observations at any given time step on average. Fig.~\ref{fig:STEI} (Right) further reveals frequent merging, with an average Merge Factor of 0.33 implying that 3 out of 10 individuals are involved in a merge event at each time step on average. Together, these metrics clearly demonstrate the challenge in resolving individual footprints from point clouds, underscoring the need for a new paradigm.

\textbf{Remark:} We note that only merge and occlusion occur frequently, but each such event can persist for extended periods, resulting in prolonged periods of observability loss, as we shall quantify in Section~\ref{sec:exp_val}. 


\section{Proposed Methodology: Macro to Micro Crowd Untangling}
\label{sec:proposed_solution}
In this section, we present a novel framework for untangling dense crowd observations and recovering the spatial behaviors of individuals in crowds exhibiting strong spatio-temporal entanglement. To this end, we adopt  a two-scale perspective: at the macroscopic level, we extract crowd semantics that capture aggregate spatial usage patterns from radar point clouds. At the microscopic level, 
we shift to reconstruct individual spatial footprints guided by these learned spatial semantics. As discussed in Sec. \ref{sec:introduction}, spatio-temporal entanglement leads to recurrent merged observations and intermittent visibility in the radar point clouds, resulting in systematic loss of information regarding individual movement within the crowd. To address this, we propose a reasoning mechanism that integrates learned spatial semantics with models of merging and blockage to systematically generate, score, and propagate hypotheses, resolving ambiguities in entangled observations to reveal individual spatial behaviors.

\subsection{Learning Collective Spatial Semantics -- A Macroscopic Perspective}
\label{subsec:flow_field_estimation}
\vspace{-3pt}

We begin by extracting a macroscopic representation of crowd motion from the temporally evolving radar point clouds $\mathcal{P}(t)$. Although individual radar detections are sparse and intermittent, aggregate observations reveal persistent patterns of how space is collectively used by the crowd. 
Building on prior work in radar-based flow field estimation \cite{pallaprolu2025mmflux, Zhang2024Sensors}, we capture such patterns through a continuous spatial flow field $\hat{\mathbf{v}}(\mathbf{x}): \mathbb{R}^2 \rightarrow \mathbb{R}^2$, which encodes the dominant motion tendencies at each location $\mathbf{x}$ in the FOV.

To estimate this field from radar data, we compute local motion statistics directly from the point clouds. Each point cloud is first embedded into a binary spatial occupancy map, after which local displacements are estimated over short temporal windows using classical optical flow techniques \cite{lucas1981iterative}. Owing to the inherent sparsity of radar point clouds, these initial motion estimates are noisy and inconsistent. However, true collective motion manifests as consistent directional structure over time. We therefore analyze the angular distribution of flow vectors at each location $\mathbf{x}$ and retain only those regions exhibiting statistically significant deviation from a uniform distribution $\mathcal{U}(0, 2\pi)$, via a Kolmogorov-Smirnov test \cite{pallaprolu2025mmflux}. The retained vectors are averaged over time to obtain an initial estimate of the flow field. Finally, spatial smoothing is applied to enforce coherence across neighboring regions, yielding a continuous field $\hat{\mathbf{v}}(\mathbf{x})$ (illustrated in Fig. \ref{fig:pipeline_flowchart}) together with an associated covariance $\hat\Sigma(\mathbf{x})$ that captures uncertainty in the estimated motion tendencies. We note that only a brief observation period is needed in practice to compute the semantics (e.g., a few seconds).

The learned macroscopic semantics $\hat{\mathbf{v}}(\mathbf{x})$ serve as our baseline for predicting how individuals move through the space, as we discuss next. Specifically, we represent the state of the $i^{\text{th}}$ individual at time $t$ as $\mathbf{s}^i_t = [\mathbf{x}^i_t\ \ \mathbf{v}^i_t]^\top$, which captures 2D position and velocity of the individual. Classical models of pedestrian motion over short temporal windows often assume constant-velocity dynamics \cite{scholler2020constant}. In our setting, however, individual motion is shaped by the macroscopic spatial semantics $\hat{\mathbf{v}}(\mathbf{x})$ learned from the collective evolution of the radar point clouds. More specifically, the predicted state evolves according to
\setlength{\abovedisplayskip}{2pt}
\begin{align}
\label{eq:motion_model}
    \mathbf{s}^i_{t} &= \mathbf{F}_\text{sem}(\mathbf{s}^i_{t-1}; \mathbf{\hat{v}}) + \mathbf{w}_t,\ w_t \sim \mathcal{N}(0,\mathbf{Q}), \text{  where}
\end{align}
\begin{align}
\mathbf{F}_\text{sem}(\mathbf{s}^i_{t-1}; \hat{\mathbf{v}}) =
\begin{bmatrix}
\mathbf{x}^i_{t-1} + \Delta t\,\mathbf{v}^i_{t-1} \\
\hat{\mathbf{v}}(\mathbf{x}^i_{t-1} + \Delta t\,\mathbf{v}^i_{t-1})
\end{bmatrix}.
\end{align}
\setlength{\belowdisplayskip}{2pt}
The operator $\mathbf{F}_\text{sem}(\cdot, \hat{\mathbf{v}})$ defines a semantic-guided motion model, in which the macroscopic flow field $\hat{\mathbf{v}}(\mathbf{x})$ acts as a location-dependent motion cue, coupling microscopic individual motion with the dominant patterns of collective movement captured from the radar point clouds. However, reconciling these predictions with radar observations becomes considerably challenging in dense crowds due to frequent observability loss. As a result, semantic motion guidance alone is insufficient for reliable footprint inference. We therefore next examine the physical mechanisms underlying observability loss and develop a novel and systematic reasoning framework that jointly incorporates the learned semantics with principled models of merging and occlusion.
\vspace{-10pt}

\subsection{Modeling Observability Loss under Spatio-Temporal Entanglement}
\label{subsec:loss_of_obs_ste}
\begin{figure}
    \centering
    \includegraphics[width=\linewidth]{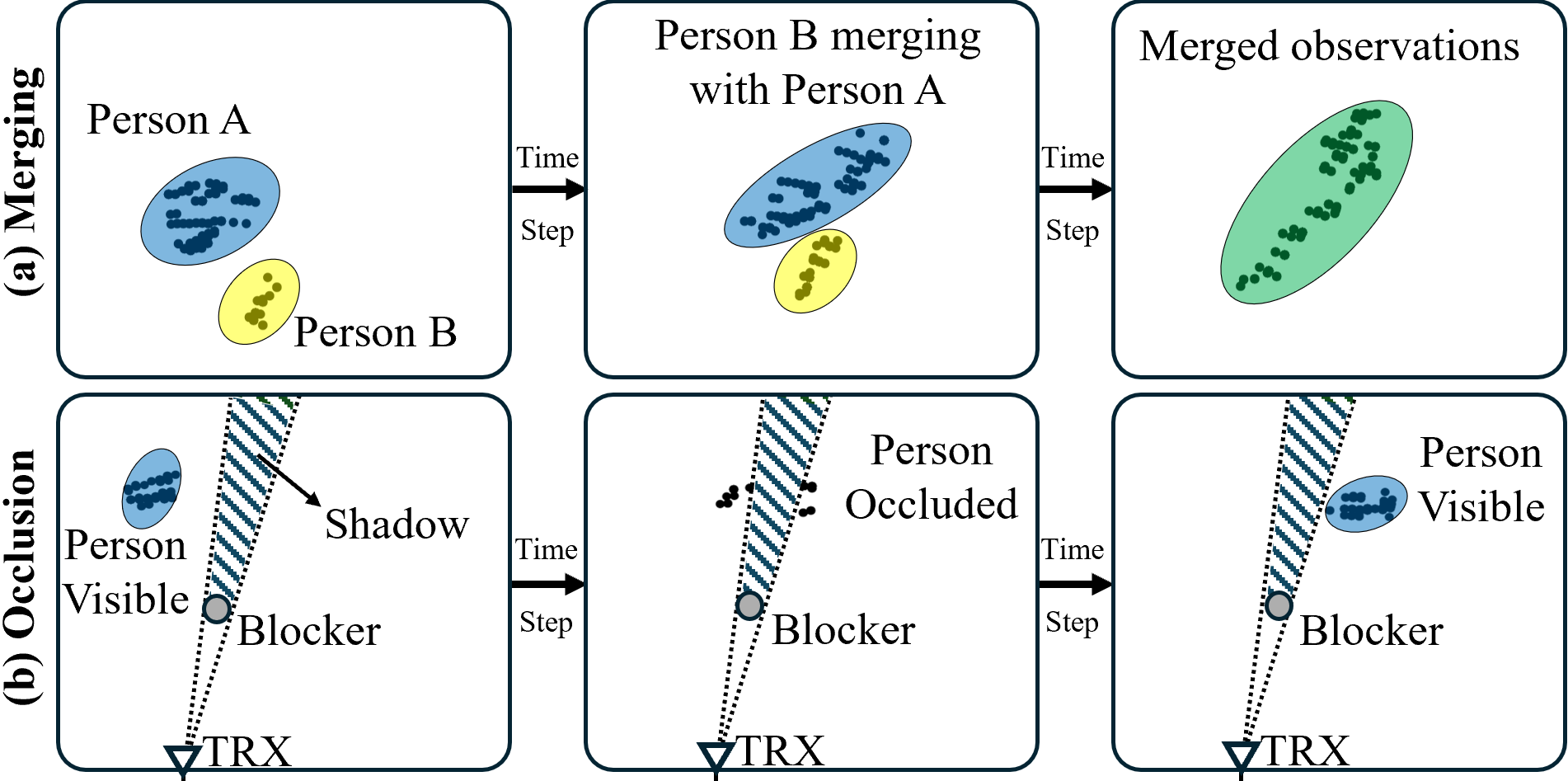}
    \vspace{-15pt}
    \caption{Loss of observability due to spatio-temporal entanglement. Ellipses denote clusters of radar returns. (a) As two spatially separated individuals move into close proximity, their measurements merge into a single cluster, causing loss of individual observability. 
    (b) Occlusion by a disc-shaped blocker closer to the TRX leads to sparse returns. See color PDF for best viewing.}
    \vspace{-20pt}
    \label{fig:loss_of_obs}
\end{figure}

\vspace{-2pt}
Limited angular resolution and the quasi-optical propagation characteristics of mmWave radiation~\cite{rangan2014millimeter} result in observations that no longer maintain a simple one-to-one correspondence with individual targets. Two physical phenomena dominate this loss of observability:

\textbf{(a) Merging:} When multiple individuals move in close spatial proximity, their reflected radar returns may appear as a single clustered detection. In such cases, the radar produces one observation corresponding to more than one individual, resulting in a many-to-one relationship between states and observations. Fig. \ref{fig:loss_of_obs} (a) illustrates merging in real radar data.

\textbf{(b) Occlusion:} An individual located farther from the radar may be partially or completely blocked by another individual closer to the radar (Fig. \ref{fig:loss_of_obs} (b)). Specifically, due to human body's high penetration loss at mmWave frequencies~\cite{slezak2022measurement}, an occluding individual can eliminate radar returns of its shadow as established in the literature~\cite{gustafson2012characterization}, resulting in total loss of observability. 

These phenomena are recurring consequences of strongly entangled crowd motion, thereby introducing ambiguity in the relationship between individuals and the radar observations they generate. As a result, the observation set at any instant may represent a superposition of multiple possible physical explanations arising from merging and occlusion events. In the next section, we introduce a novel microscopic reasoning framework that treats observability loss as a principled signal for untangling individual footprints through joint reasoning over merging, occlusion, and learned semantics.
\vspace{-10pt}

\subsection{Semantic-Guided Reasoning under Information Loss -- A Microscopic Perspective}
\label{subsec:multiple_hypothesis_tracking}
\vspace{-2pt}
Resolving spatio-temporal entanglement in dense crowds requires a systematic and integrated reasoning framework that jointly accounts for the learned crowd semantics, observability loss due to merging and occlusion, and the ambiguous radar observations arising from these phenomena. On the one hand, the learned semantic field $\hat{\mathbf{v}}(\mathbf{x})$ in Sec. \ref{subsec:flow_field_estimation} provides a macroscopic perspective of spatial usage. On the other hand, the radar point clouds $\mathcal{P}(t)$ provide sparse, entangled evidence of the instantaneous spatial configuration of the crowd. When merging and occlusion occur, however, these observations no longer admit a unique explanation in terms of individual footprints, giving rise to multiple plausible interpretations of the same set of observations. The key challenge is therefore to methodically integrate semantic motion dynamics with radar evidence in a way that can reason over these competing explanations. We tackle this challenge by treating competing interpretations explicitly as hypotheses, propagating them forward in time and resolving ambiguity as new radar evidence accumulates. We next start by formalizing the radar observation model, describe how predicted individual states are reconciled with radar measurements. 

Radar observations derived from the clustered point clouds are modeled through the nonlinear measurement function $\mathbf{z}^i_t = \hat{\mathbf{z}}^i_t + \eta_t = h(\mathbf{s}^i_t) + \eta_t, \eta_t \sim \mathcal{N}(0,\mathbf{R}),$ where $\mathbf{z}^i_t$ denotes the radar observation corresponding to the $i^{\text{th}}$ individual at time $t$, and $h(\cdot)$ maps the underlying state to the noise-free observation $h(\mathbf{s}^i_t) = \hat{\mathbf{z}}^i_t = [\mathbf{x}^i_t, \ \ v^i_{r, t}]^\top = [\mathbf{x}^i_t, \ \ (\mathbf{x}^i_t\cdot \mathbf{v}^i_t)/||\mathbf{x}^i_t||]^\top.$

At each time step $t$, our reasoning framework represents the state of the $i^{\text{th}}$ individual as a Gaussian belief $\mathcal{N}(\hat{\mathbf{s}}^i_{t|t-1}, \mathbf{P}^i_{t|t-1})$, where $\hat{\mathbf{s}}^i_{t|t-1}$ denotes the predicted state estimate and $\mathbf{P}^i_{t|t-1}$ denotes the state covariance. These predictions are obtained by propagating estimates from $t-1$ using the semantic-guided motion model described in Eq. \ref{eq:motion_model}. The radar observations at time $t$, denoted by the set $\{\mathbf{z}^j_t\}_{j=1}^{N_t}$, must then be assigned to the predicted states in order to update the inferred footprints of individuals in the crowd. 

As discussed in Sec. \ref{subsec:loss_of_obs_ste}, multiple assignments may plausibly explain the same set of measurements due to observability loss, each corresponding to a different physical explanation of how individuals generated the radar returns. We therefore formulate the disentanglement problem in terms of correspondence hypotheses that jointly explain the predicted states and the incoming observations. At every time step, hypotheses are instantiated, scored based on their consistency with both the radar measurements and the learned crowd semantics, and propagated forward in time. The resulting reasoning process forms a dynamic hypothesis tree that maintains multiple competing explanations of individual motion until sufficient evidence emerges to disambiguate them, with ambiguities typically resolving within a short temporal horizon as new observations arrive. 

More specifically, we consider three canonical classes of correspondence hypotheses within this framework: (i) a baseline hypothesis in which each observation corresponds to a single individual, (ii) merge hypotheses in which multiple individuals jointly generate a single observation, and (iii) occlusion hypotheses in which an individual produces no observable return due to blockage. We next describe how each hypothesis class is constructed and evaluated within our proposed semantic-guided reasoning framework.

\subsubsection{Baseline Hypothesis}
\label{subsubsec:baseline_hypothesis}
Our baseline correspondence hypothesis assumes that each radar observation arises from a single individual and that each individual generates at most one observation at a given time step. We instantiate this hypothesis,  denoted by $\mathcal{B}_t$, by constructing a cost matrix $C_{ij,t}$ using the Euclidean distance between predicted positions of individuals $\{\hat{\mathbf{s}}^i_{t|t-1}\}$ and observations $\{\mathbf{z}^j_t\}$. A minimum-cost one-to-one assignment is then computed using the Hungarian algorithm~\cite{kuhn1955hungarian}, following the global nearest neighbor (GNN) strategy \cite{blackman1999design}. An assignment $(i,j)$ is accepted only if the corresponding distance cost satisfies $C_{ij,t} \leq \epsilon_B$.

For each accepted assignment $(i,j)\in\mathcal{B}_t$, the predicted belief $\mathcal{N}(\hat{\mathbf{s}}^i_{t|t-1}, \mathbf{P}^i_{t|t-1})$ is corrected using the radar observation $\mathbf{z}^j_t$ through an Extended Kalman Filter (EKF) update \cite{bar2001estimation}, yielding the posterior estimate $\mathcal{N}(\hat{\mathbf{s}}^i_{t|t}, \mathbf{P}^i_{t|t})$.
Importantly, this update integrates the semantic-guided motion prediction with the incoming radar observation to describe the belief over the state of the individual in the crowd. We then assign a score to the updated state estimate of the $i^{\text{th}}$ individual in $\mathcal{B}_t$ as $\rho(i, t) = -\log(|\mathbf{P}^i_{t|t}|)$, where $|\cdot|$ is the determinant. The overall score of hypothesis $\mathcal{B}_t$ is then defined as
\setlength{\abovedisplayskip}{2pt}
\begin{align}
\label{eq:baseline_assoc_cost}
    \Upsilon(\mathcal{B}_t) = \sum_{(i, j)\in\mathcal{B}_t}\rho(i, t) = -\sum_{(i, j)\in\mathcal{B}_t}\log(|\mathbf{P}^i_{t|t}|).
\end{align}
\setlength{\belowdisplayskip}{2pt}
The choice of $\rho(i, t)$ is motivated by its equivalence (up to an additive constant) to the differential entropy~\cite{cai2015law} of the multivariate Gaussian posterior estimate $\mathcal{N}(\hat{\mathbf{s}}^i_{t|t},\mathbf{P}^i_{t|t})$. 
Since the covariance evolves through both prediction and measurement updates, the score naturally reflects the accumulation or absence of observational evidence over time.

We incorporate spatial crowd semantics ($\hat{\mathbf{v}}(\mathbf{x}), \hat\Sigma(\mathbf{x})$), introduced in Sec.~\ref{subsec:flow_field_estimation}, directly into the hypothesis scoring process to favor motion patterns consistent with the macroscopic space usage patterns. Specifically, we modify $\rho(i,t)$ to penalize deviations of the estimated velocity from the expected velocity informed by $\hat{\mathbf{v}}(\mathbf{x})$ at the estimated position of the $i^{\text{th}}$ individual. Let the posterior state estimate be partitioned as $\hat{\mathbf{s}}^i_{t|t} = [\hat{\mathbf{x}}^{i}_{t|t}\ \ \hat{\mathbf{v}}^{i}_{t|t}]^\top$, where $\hat{\mathbf{x}}^{i}_{t|t}$ and $\hat{\mathbf{v}}^{i}_{t|t}$ denote the position and velocity components, respectively. We define the semantic-guided score as
\setlength{\abovedisplayskip}{2pt}
\begin{align}
\label{eq:baseline_single_track_ff_cost}
    \rho(i, t) &= -(\log(|\mathbf{P}^i_{t|t}|) + \log(|\hat\Sigma(\hat{\mathbf{x}}^{i}_{t|t})|) + \xi(i, t)), \text{where} \\
\xi(i, t) &= (\hat{\mathbf{v}}(\hat{\mathbf{x}}^{i}_{t|t}) - \hat{\mathbf{v}}^{i}_{t|t})\ (\hat\Sigma(\hat{\mathbf{x}}^{i}_{t|t}))^{-1}\ (\hat{\mathbf{v}}(\hat{\mathbf{x}}^{i}_{t|t}) - \hat{\mathbf{v}}^{i}_{t|t})^\top. \nonumber
\end{align}
\setlength{\belowdisplayskip}{2pt}
Consequently, the modified score favors associations that are both statistically confident (small $|\mathbf{P}^i_{t|t}|$) and dynamically consistent with the underlying spatial usage patterns.  

\textbf{Remark:} While we refer to this hypothesis as our baseline, it does not correspond to any prior method, as it integrates our learned semantics with motion model predictions.

While the baseline hypothesis enforces a one-to-one correspondence between individuals and observations, this assumption may be violated under spatio-temporal entanglement in dense crowds (see Sec. \ref{subsec:loss_of_obs_ste}). We next introduce the merge hypothesis class to explicitly model the possibility that multiple predicted states correspond to a single observation.
\subsubsection{Merge Hypothesis}
\label{subsubsec:merge_hypothesis} 
We introduce merge hypotheses to capture cases where two or more states $\hat{\mathbf{s}}^{i_1}_{t|t-1},\hat{\mathbf{s}}^{i_2}_{t|t-1} \cdots$ yield Euclidean distance costs to the same observation $\mathbf{z}^j_t$ within a threshold $C_{i_kj, t}\leq\epsilon_M$. In such cases, we interpret the event as a potential merge and we form a clique of individual states as the merge hypothesis $\mathcal{M}_{j, t} = \{i_1, i_2, \cdots\}$. 
Importantly, the individual states are not fused into a single estimate, instead each state in the clique is corrected independently using the shared observation through its own EKF update, preserving the separability of each individual through the merge event. 
The overall score of the corrected states in $\mathcal{M}_{j, t}$ is
\setlength{\abovedisplayskip}{2pt}
\begin{align}
    \Upsilon(\mathcal{M}_{j, t}) = \sum_{i\in\mathcal{M}_{j, t}}\rho(i, t)/|\mathcal{M}_{j, t}| = -\!\!\sum_{i\in\mathcal{M}_{j, t}}\log(|\mathbf{P}^i_{t|t}|)/|\mathcal{M}_{j, t}|, \nonumber
\end{align}
\setlength{\belowdisplayskip}{2pt}
where $|\cdot|$ denotes cardinality. We incorporate the semantic-guided state score $\rho(i,t)$ (defined in Eq.~\ref{eq:baseline_single_track_ff_cost}) while evaluating $\Upsilon(\mathcal{M}_{j, t})$. \todo{This unified formulation jointly evaluates observation consistency and compatibility with macroscopic semantics. Although a merged observation may attract multiple predictions toward the same location, the state estimates must also remain consistent with the underlying spatial semantics. This additional constraint allows the framework to recover spatially distinct footprints, even during prolonged merge events, by balancing observation evidence with the expected macroscopic crowd distribution.}


While the baseline and merge hypotheses explain radar observations that may correspond to one or more individuals, some predicted states may remain unsupported by any radar return under either hypothesis. As shown in Fig. \ref{fig:loss_of_obs} (b), this typically arises when an individual becomes temporarily occluded while traversing through the shadow cast by a blocker with respect to the radar line of sight. We next introduce an occlusion hypothesis class to explicitly model such events and provide a principled explanation for these observational outages in entangled radar observations.

\subsubsection{Occlusion Hypothesis} 
\label{subsubsec:occ_hypothesis}
Predicted individual states that admit neither a baseline nor a merge correspondence with a radar observation are treated under the occlusion hypothesis $\mathcal{O}_t$, which we define as the set of such unsupported states at time $t$. Since no radar observation is available, the predicted belief $\mathcal{N}(\hat{\mathbf{s}}^i_{t|t-1}, \mathbf{P}^i_{t|t-1})$ of an occluded individual $i$ cannot be enhanced through an EKF update and therefore evolves purely under the semantic-guided motion model of Eq. \ref{eq:motion_model}. 

As discussed in Sec. \ref{subsec:loss_of_obs_ste}, an individual becomes occluded when it lies within the geometrical shadow cast by another individual located closer to the radar along the line of sight. We define the shadow region of the $k^{\text{th}}$ individual at time $t$ in polar coordinates as $\Omega_{k, t} = \{(r, \theta)\mid r > r_{k, t}, |\theta - \theta_{k, t}| < \arcsin(d_h/r_{k, t})\}$, where $d_h$ denotes the effective radius of an individual~\cite{biacromialCDC}, and $(r_{k, t}, \theta_{k, t})$ is the predicted position of the $k^{\text{th}}$ individual. The $i^{\text{th}}$ individual is said to be occluded by the $k^{\text{th}}$ individual at time step $t$ if $(r_{i, t}, \theta_{i, t}) \in \Omega_{k, t}$. 

We evaluate occlusion probabilistically, and we define the relative polar state between individuals $i$ and $k$ as $\boldsymbol{z}_t = [z_{r,t}\ \ z_{\theta, t}]^T\sim \mathcal{N}(\boldsymbol{\mu}_{z,t}, \boldsymbol{\Sigma}_{z,t})$, where $\boldsymbol{\mu}_{z,t} = \boldsymbol{\mu}_{r\theta,i,t} - \boldsymbol{\mu}_{r\theta,k,t},\ 
\boldsymbol{\Sigma}_{z,t} = \boldsymbol{\Sigma}_{r\theta,i,t} + \boldsymbol{\Sigma}_{r\theta,k,t},$ and $\boldsymbol{\mu}_{r\theta,t}$ and $\boldsymbol{\Sigma}_{r\theta,t}$ are obtained by linearizing the Cartesian-to-polar transform about the mean position~\cite{thrun2005probabilistic}. The probability that individual $i$ lies within $\Omega_{k,t}$ is then given by $\phi_{ik,t} = \mathbb{P}(z_{r,t} > 0, |z_{\theta,t}| <\arcsin(d_h/r_{k,t}))$, computed via the bivariate Gaussian CDF. Thus, we define the worst-case occlusion likelihood as $\phi_{i,t}^\star = \max_{k \neq i} \phi_{ik,t}$, reflecting the most geometrically plausible occluder for individual $i$. In total, the score of $\mathcal{O}_t$ is given by
\begin{align}
\label{eq:occlusion_assoc_cost}
    \Upsilon(\mathcal{O}_t) = \sum_{i\in\mathcal{O}_t}\rho_\text{occ}(i, t) = -\sum_{i\in\mathcal{O}_t}\log(|\mathbf{P}^i_{t|t-1}| \textcolor{red}{/} \phi_{i, t}^\star),
\end{align}
where the posterior covariance factor reflects prediction-only uncertainty. Since prediction-only propagation inflates $|\mathbf{P}^i_{t|t-1}|$ \cite{thrun2005probabilistic}, predictions that leave individuals unobserved for extended periods are therefore assigned lower scores. Furthermore, if no other individual is positioned to explain the absence of $i$'s observation, $\phi_{i,t}^\star$ remains small, penalizing states whose disappearance lacks a geometric explanation.

\textbf{Remark:} Aside from occlusion, observations that are not assigned to any state under either baseline or merge hypotheses are treated as radar clutter. We define the clutter hypothesis $\mathcal{C}_t$ as the set of all unassociated observations $\mathbf{z}^j_t$ at time $t$. Clutter is modeled as a spatially uniform Poisson point process~\cite{bar2001estimation}, with $\Upsilon(\mathcal{C}_t) = |\mathcal{C}_t|\log(\lambda_{c, t}/A)$, where $\lambda_{c,t}$ denotes the expected number of clutter observations per time step within the FOV of area $A$.


With the baseline, merge, occlusion, and clutter hypotheses defined, we obtain a complete set of mutually exclusive explanations for all predicted states and observations at time $t$, enabling consistent comparison within a unified scoring framework. We next describe how these local hypotheses are aggregated within our semantic-guided reasoning framework, and how the resulting hypothesis tree is managed and pruned to maintain computational tractability over time.

\begin{figure*}
    \centering
    \includegraphics[width=\linewidth]{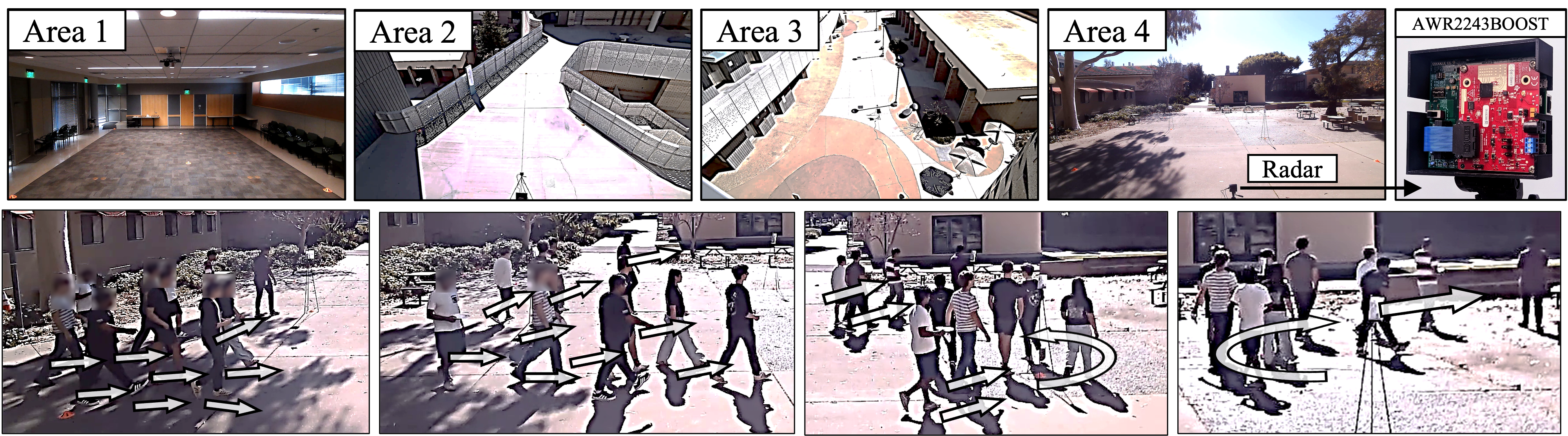}
    \caption{Experimental areas and sample evolution of crowds considered in this work. (Top) Experimental areas include an indoor multipath-rich auditorium (Area 1), a narrow elevated walkway bounded by strong metallic reflectors (Area 2), and outdoor spaces with heterogeneous obstructions such as seating, foliage, and nearby pathways (Areas 3 and 4). (Bottom) Sample video frames from one of our experiments, illustrating the crowd density and interaction complexity characteristic of our evaluation scenarios. See color PDF for best viewing.}
    \vspace{-20pt}
    \label{fig:exp_areas}
\end{figure*}

\subsubsection{Multi-Hypothesis Reasoning}
\label{subsubsec:hyp_management}
We next describe the temporal evolution of the hypothesis tree. Tree initialization can be implemented in different ways depending on how individuals enter the FOV. In our experiments, it follows the procedures described in Sec.~\ref{subsec:exp_res}. For the present discussion, we assume an initialized hypothesis tree. Specifically, consider the case that at time $t_0$ a set of active predicted states has been established and the recursion begins at $t=t_1$ by forming an initial set of hypotheses $\mathcal{H}_{t_1} = \{\mathcal{B}_{t_1}\}$, consisting of the baseline correspondences with observations $\mathbf{z}_{t_1}^{j}$. States that do not receive a baseline assignment give rise to local occlusion hypotheses $\mathcal{O}_{t_1}$, while observations without corresponding baseline assignments form local clutter hypotheses $\mathcal{C}_{t_1}$. In parallel, we test for clique formations among $\mathbf{z}_{t_1}^{j}$ and construct local merge hypotheses $\mathcal{M}_{j,t_1}$ whenever the conditions in Sec.~\ref{subsubsec:merge_hypothesis} are satisfied. The global hypothesis set $\mathcal{H}_{t_1}$ is then augmented with feasible local hypotheses.

More generally, let $\mathcal{H}_{t_{k-1}}$ denote the set of feasible hypotheses at time $t_{k-1}$. For each parent hypothesis $h_p \in \mathcal{H}_{t_{k-1}}$ at a general time $t_k$, we construct the set of feasible local hypotheses $\mathcal{L}_{t_k} = \{\mathcal{M}_{j,t_k}\}_{j=1}^{J_{t_k}} \cup \{\mathcal{B}_{t_k}, \mathcal{O}_{t_k}, \mathcal{C}_{t_k}\}.$ Each hypothesis $\ell \in \mathcal{L}_{t_k}$ generates a child $h\in\mathcal{H}_{t_k}$ by extending the parent $h = h_p \oplus \ell$, where $\oplus$ denotes a concatenation operation. We subsequently perform the EKF update in a priority-ordered manner. Individuals with predicted states participating in merge hypotheses are updated first, followed by states under the baseline hypothesis, restricted to individuals not already updated under a merge hypothesis. Finally, the predicted belief itself serves as the update for states under occlusion hypotheses, which by definition lack a corresponding observation. This ordering guarantees that each track is updated at most once per time step, after which the set $\mathcal{H}_{t_k}$ is scored.



We next define the cumulative (global) score at time $t=t_k$ for each hypothesis $h \in \mathcal{H}_{t_k}$ recursively. Let $h_p \in \mathcal{H}_{t_{k-1}}$ denote the unique parent from which $h$ is generated by instantiating $\ell$ at time $t_k$. The accumulated score is given by $S(h) = S(h_p) + \Upsilon(\ell)$. This recursive formulation ensures that each global hypothesis maintains the cumulative sum of the local hypothesis scores along its ancestral branch. To maintain computational tractability, we allow the hypothesis set $\mathcal{H}_{t_k}$ to expand for $t_p$ time steps before performing pruning, at which point only the top $P$ hypotheses (ranked by cumulative score $S(\cdot)$) are retained and the remainder discarded. The parameter $t_p$ controls the temporal depth over which ambiguity is preserved prior to pruning, whereas $P$ determines the maximum number of competing hypotheses that are allowed to survive at each pruning stage. Thus, $t_p$ regulates how long alternative explanations are explored over time, while $P$ regulates the branching width of the tree, trading off hypothesis diversity against computational cost. 

We next present experimental validation to demonstrate the effectiveness of the proposed methodology in untangling individual footprints.
\section{Experimental Validation} 
\label{sec:exp_val}
We next present an extensive experimental evaluation of our proposed foundation for recovering individual footprints in spatio-temporally entangled crowds. We begin by describing our mmWave radar platform and four experimental areas selected in natural, everyday environments, along with our camera-based pipeline for generating ground-truth individual footprints. We then describe the six crowd topologies designed to emulate representative natural crowd behaviors, spanning 18 experiments with crowd sizes of up to 10 individuals.\footnote{This research has been reviewed and approved by our Institutional Review Board (IRB) committee.}  We then present untangling performance across all experiments, demonstrating strong structural consistency with the ground truth. The recovered footprints further enable accurate zonal analysis of pedestrian space usage. Finally, we present an ablation study highlighting the contribution of key components of our framework.






\subsection{Experimental Setup}
We evaluate our framework using the TI AWR2243BOOST off-the-shelf mmWave FMCW radar module~\cite{awr2243boostug}, operating in the 76–81 GHz automotive band. The radar is initialized with a start frequency of $f_0 = 76~\text{GHz}$ and transmits chirps spanning a bandwidth of $B = 4.24~\text{GHz}$. We enable all $N_\text{TX} = 3$ transmit antennas and treat the virtual MIMO array as a linear horizontal aperture due to limited elevation resolution. The raw ADC samples are captured through a DCA1000EVM data capture card~\cite{dca1000evm} and streamed to a host machine over Ethernet for point cloud generation.

\begin{figure*}[t!]
    \centering
    \includegraphics[width=\linewidth]{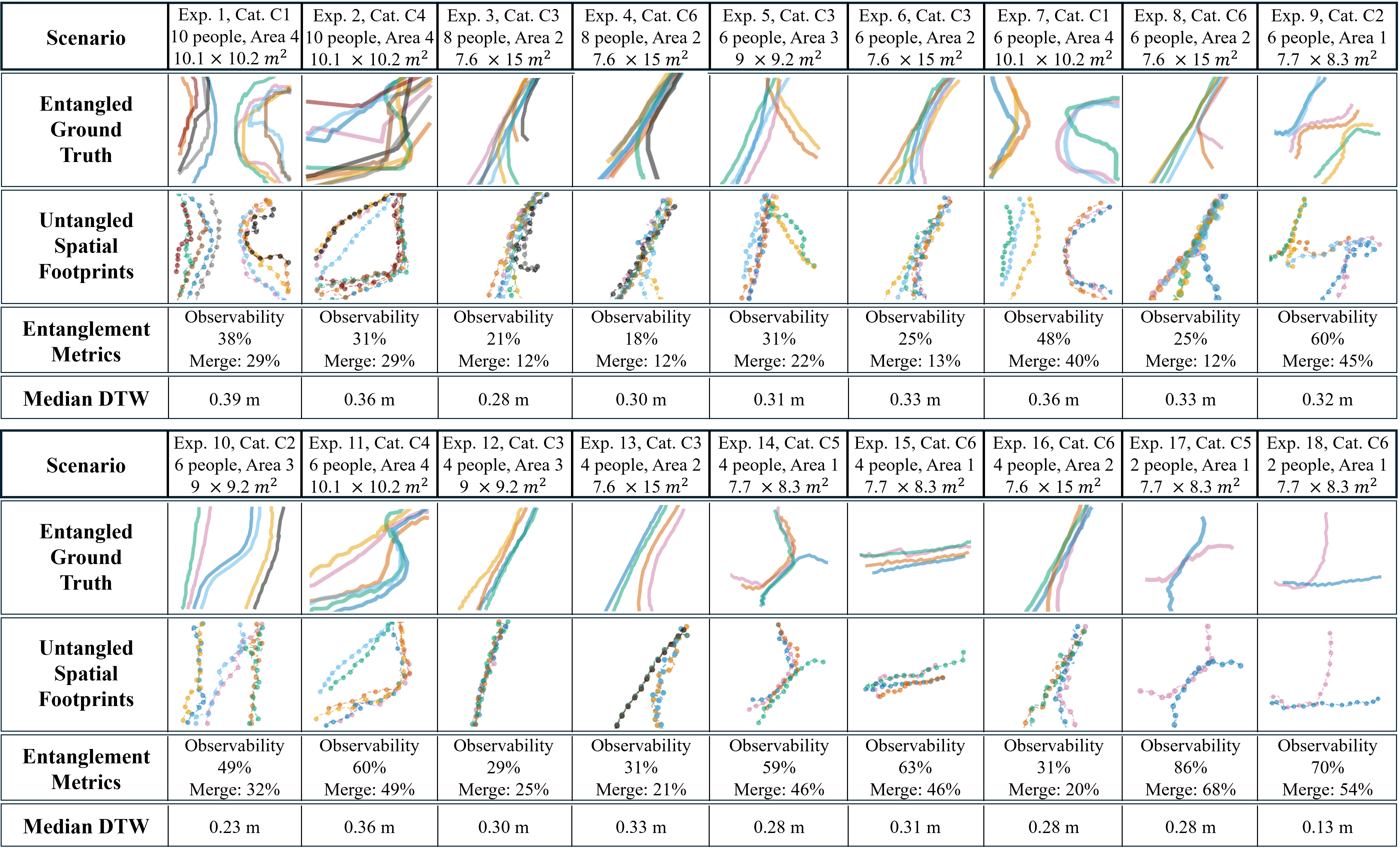}
    \caption{Untangling performance of our proposed methodology. As seen, our results exhibit strong visual and structural agreement with the ground-truth.
    We further achieve an average median DTW of 0.31 m across all 18 experiments, demonstrating robust individual footprint recovery across diverse crowd topologies and varying levels of spatio-temporal entanglement. See color PDF for best viewing.}
    \vspace{-18pt}
    \label{fig:all_results_table}
\end{figure*}

\subsubsection{Experimental Areas} We conduct experiments across four distinct environments, illustrated in Fig.~\ref{fig:exp_areas} (Top), which reflect naturally occurring scene geometries and diverse propagation conditions. Area 1 is a 7.79 $\times$ 8.39 m indoor auditorium with an open floor plan bounded by reflective walls and a ceiling, representing a multipath-rich environment. Area 2 is a 15 m footbridge enclosed by metallic railings, whose narrow entryway naturally compresses pedestrian flow into tightly coupled motion. Areas 3 and 4 are open outdoor spaces of 9 $\times$ 9.27 m and 10.1 $\times$ 10.29 m respectively, surrounded by mixed obstructions including seating, foliage, and a nearby bike path. Fig.~\ref{fig:exp_areas} (Bottom) shows a sample crowd evolution from Area 4, illustrating how pedestrian behaviors, e.g., blocking and merging, naturally lead to spatio-temporal entanglement, characteristic of the considered scenarios.

\subsubsection{Ground Truth Generation} Our setup includes a Lenovo FHD 300 camera~\cite{lenovo300webcam} mounted overhead, recording a monocular video stream concurrently with radar measurements to generate ground-truth individual footprints. We extract per-person motion footprints from the video stream using HybridSORT~\cite{yang2024hybrid}, which robustly integrates motion and appearance cues. To recover metric footprints from monocular observations, we leverage knowledge of the experimental region and apply Inverse Perspective Mapping (IPM)~\cite{hartley2004multiple} to the bottom-center of each detected bounding box. 

\subsubsection{System Parameters} 
Within our multi-hypothesis reasoning module, $\epsilon_M$ = 0.35 m defines the merge threshold below which two individuals appear as a single observation, and $\epsilon_B = 2\epsilon_M = 0.7$ m gates baseline hypotheses. To ensure computational tractability, the hypothesis tree is pruned every $t_p = 5$ frames, retaining the top $P = 5$ hypotheses ranked by cumulative likelihood. Process noise follows a white-acceleration model~\cite{sarkka2013bayesian} with $\sigma_a = 1.5 \text{m/s}^2$, and measurement noise parameters are set to $\sigma_x = \delta_r\cos(\delta_\theta)/2, \sigma_y = \delta_r\sin(\delta_\theta)/2$, and $\sigma_{v_r} = \delta_{v_r}/2$, where $\delta_r, \delta_\theta, \delta_{v_r}$ are the range, angle, and Doppler resolutions of the radar.



\vspace{-8pt}
\subsection{Experimental Results}
\vspace{-3pt}
\label{subsec:exp_res}
We begin by describing the six crowd topologies and the extent of spatio-temporal entanglement they exhibit. We then present a comprehensive evaluation of untangling performance across all 18 experiments, followed by zonal analysis of individual space usage patterns and ablation studies.

\subsubsection{Categories of Crowd Behaviors} We conduct 18 experiments across four areas with crowd sizes of up to (and including) 10 individuals, organized into six categories (C1–C6) that capture distinct forms of spatio-temporal coupling. Fig.~\ref{fig:all_results_table} provides representative examples of each category, showing individual spatial footprints recovered by our framework below the camera-derived ground truth paths. Before presenting the results, we first outline the diverse categories of crowd behaviors considered.


C1 captures opposing pedestrian streams entering from opposite ends of the FOV, converging near the center, and continuing toward opposite exits -- a particularly challenging configuration producing dense spatio-temporal coupling both within and across streams. Such patterns frequently arise in open plazas or transit concourses where pedestrians traveling between different destinations cross paths near a central walkway. C2 captures intermittent group-switching, where individuals dynamically shift between nearby groups, producing transient merges and splits as the crowd evolves. C3 represents destination-driven divergence, where individuals enter the scene together and gradually deviate toward different final directions, as observed at stadium or transit station entrances. C4 emulates highly interactive environments such as poster sessions or exhibitions, where individuals temporarily slow down around points of interest before continuing through the space; a sample temporal evolution of a C4 configuration for a 10 person crowd in Area 4 is illustrated in Fig.~\ref{fig:exp_areas} (Bottom). C5 and C6 capture structured merging and splitting behaviors commonly observed in constrained pedestrian corridors. In C5, pedestrian streams that are initially separate temporarily merge into a shared flow before diverging toward different exits, whereas in C6 individuals move together initially and then choose if they want to diverge at a well-defined 
region, as seen in 
transit hubs.

\begin{figure*}[t!]
    \centering
    \includegraphics[width=\linewidth]
{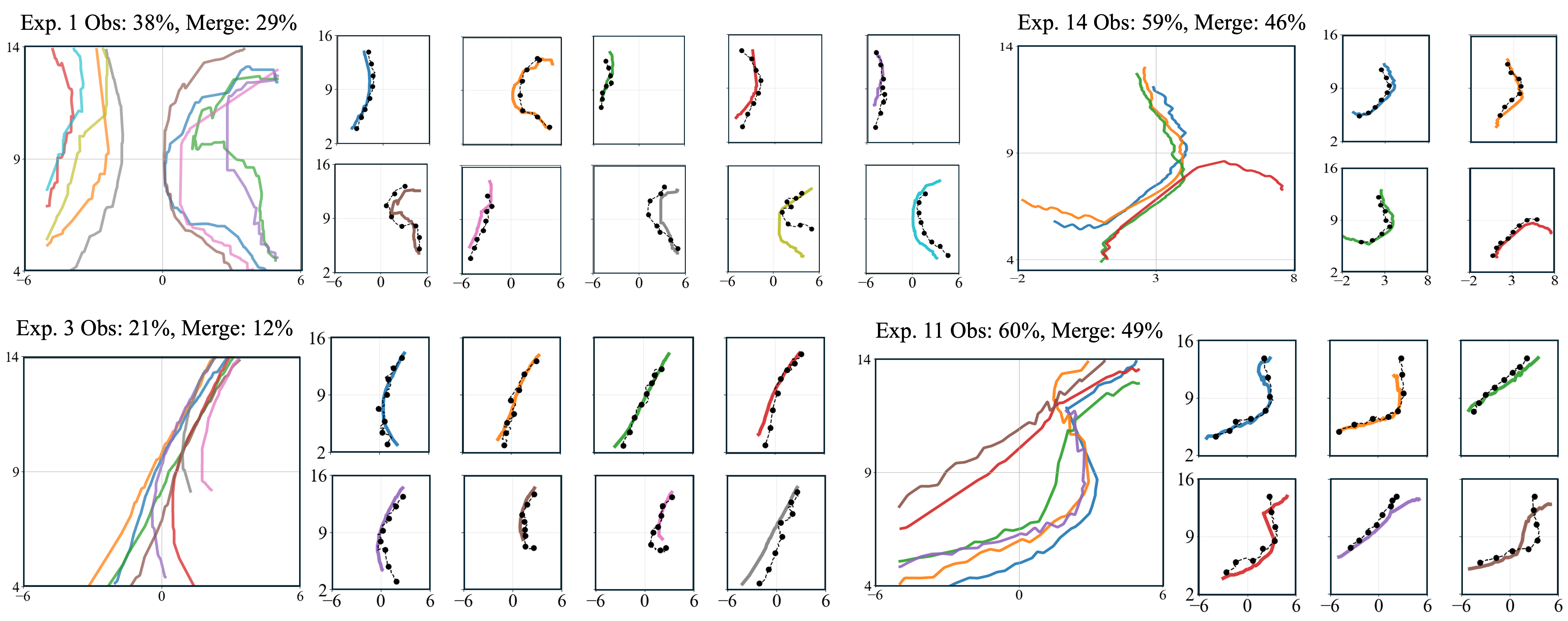}
\vspace{-22pt}
    \caption{Zooming in on individual footprint recovery performance for Experiments 1, 3, 11, and 14 of Fig.~\ref{fig:all_results_table}. Each recovered footprint (dotted line) is overlaid against its camera-derived ground truth counterpart, demonstrating strong spatial correspondence across varying crowd sizes and spatial usage patterns. See color PDF for best viewing.}
    \label{fig:sample_untangling_results}
    \vspace{-22pt}
\end{figure*} 

In all but one configuration, individuals enter the scene from one region of the FOV and progress toward exit points on the opposing side. C1 is the exception, capturing opposing pedestrian streams entering from opposite ends of the FOV, converging near the center, and continuing toward opposite exits. While participants were briefed on the spatial layout of each topology during experimentation, they were not asked to follow any predefined routes. Instead, they moved freely within the given scenario, capturing natural motion dynamics and  inherent variability of real-world crowd behavior.


\subsubsection{Extent of Spatio-Temporal Entanglement}
In Sec.~\ref{sec:radar_obs_in_dense_crowds}, we introduced two metrics, the \textit{Observability Factor} and the \textit{Merge Factor}, to quantify the degree of crowd spatio-temporal entanglement. These metrics provide a principled characterization of the entanglement present in each scenario and are used throughout the following discussion to contextualize the experimental conditions. Fig.~\ref{fig:STEI} shows their distributions across all experiments, and Fig.~\ref{fig:all_results_table} reports the Observability and Merge Factors for each experiment. As can be seen, the crowds in our experiments exhibit a high degree of entanglement. For example, Exp.~1, involving 10 participants in Area~4, yields an Observability Factor of 38\% and a Merge Factor of 29\%. This indicates that, on average, only 38\% of occupants are individually observable by the radar at any given time, while 29\% of occupants are merged with others in the radar observations at any given time on average, highlighting the substantial ambiguity present in the measurements. Experiments 3, 4, and 6 (crowds of 8, 8, and 6 individuals in configurations C3 and C6) exhibit Observability Factors as low as 18--25\% with Merge Factors of 12--13\%, reflecting sustained periods where the majority of individuals yield very few distinct radar observations as they traverse the narrow footbridge in Area~2. On the other hand, Experiments 11, 14, and 18 exhibit relatively high Observability Factors of 60\%, 59\%, and 70\% respectively, which might suggest amenable sensing conditions. Yet their corresponding Merge Factors of 49\%, 46\%, and 54\% reveal that at least one merge event occurs nearly half the time, reflecting interaction-rich configurations where merging is the dominant source of entanglement rather than occlusion. 


In addition to the Observability and Merge Factors, we also characterize the temporal persistence of these merge events. Across all experiments, continuous merging persists for 19\% of the total experiment duration on average, with the longest uninterrupted merge lasting 82.5\% of the experiment duration in Experiment 11. Other representative cases include 48.5\% in Experiment 16 and 44.3\% in Experiment 6. Together, these metrics provide a holistic characterization of the challenge: merge events occur not only frequently, as reflected by the Merge Factor, but also remain sustained for extended periods, leading to prolonged losses in radar observability. This combination of frequent and persistent merging constitutes the primary challenge addressed by our disentanglement framework.


\subsubsection{Quantitative Metric for Untangling Performance} As we shall see later, our results exhibit strong visual and structural similarity to the ground truth. To complement this, we next introduce a quantitative comparison metric. The radar and monocular camera operate at different frame rates and observe the scene from distinct viewpoints, yielding motion footprints in different coordinate frames with asynchronous temporal sampling. Dynamic Time Warping (DTW) has been widely used in the literature to compare motion sequences obtained from heterogeneous sensors without requiring synchronized temporal samples or matched frame rates~\cite{jang2022lightweight, wan2025fuzzy}. We therefore adopt DTW as an evaluation metric, enabling meaningful comparison with the camera ground truth. Formally, given two sequences 
$\mathbf{X}=\{\mathbf{x}_i\}_{i=1}^n$ and $\mathbf{Y}=\{\mathbf{y}_j\}_{j=1}^m$, 
with $\mathbf{x}_i, \mathbf{y}_j \in \mathbb{R}^2$, the DTW distance is given by $\mathrm{DTW}(\mathbf{X},\mathbf{Y})
= \min_{\pi} \frac{1}{L}\sum_{k=1}^{L} \| \mathbf{x}_{i_k} - \mathbf{y}_{j_k} \|_2$, where $\pi=\{(i_k,j_k)\}_{k=1}^{L}$ satisfies
$(i_1,j_1 ) = (1,1), (i_L, j_L) = (n,m)$ and 
$(i_{k+1}-i_k,\, j_{k+1}-j_k) \in \{(1,0),(0,1),(1,1)\}$. 


\subsubsection{Performance of the Proposed Foundation} Fig.~\ref{fig:all_results_table} summarizes the untangling results across all 18 experiments, spanning a wide range of natural crowd behaviors that give rise to significant spatio-temporal entanglement. As the figure shows, our results exhibit strong visual and structural agreement with the ground-truth trajectories by comparing the corresponding rows. These results demonstrate that systematic reasoning that integrates principled modeling of radar observability with learned crowd semantics can effectively resolve severe observation ambiguities and reliably untangle individual footprints of a crowd. 

Similarly, our framework achieves an average median DTW of 0.31 m, with all experiments falling well within the biacromial shoulder width of 0.5 m~\cite{biacromialCDC} -- an important dimension used to characterize the spatial footprint of a person in urban planning~\cite{robal2025you}. This quantitatively confirms that recovered footprints are spatially reliable across all tested topologies. Further analysis of individual experiment results confirms that our framework performs reliably across the full spectrum of entanglement conditions characterized earlier. In experiments dominated by severe observability loss, such as Experiments 3, 4, and 6, our framework recovers individual footprints with median DTW errors of 0.28 m, 0.30 m, and 0.33 m respectively, demonstrating effective occlusion and merge hypothesis modeling under highly degraded observability. In experiments with moderate observability but substantial and sustained merging, such as Experiments 9, 11 and 14, our framework achieves median DTW errors of 0.32 m, 0.36 m and 0.28 m, respectively, underscoring the effectiveness of the proposed merge hypothesis modeling.


We note that our hypothesis tree is initialized with states derived from observations at known general entry regions (e.g., quarter-disk area of radius 1 m), as our goal is to recover space usage patterns once occupants enter an area. Our experiments further showcase three complex scenarios (Experiments 9, 14, and 15) where individuals enter as a tightly coupled clique, rendering such initialization intractable. For these 3, the total crowd size is assumed, and unobserved individuals are initialized by sampling uniformly across the entry region. These scenarios highlight promising future directions, including improved initialization and reasoning or the integration of crowd counts obtainable from door-entry sensors or existing mmWave-based counting methods.



\subsubsection{A Closer Look at Individual Footprint Recovery} 
Fig.~\ref{fig:sample_untangling_results} provides a closer look at four representative experiments (1, 3, 11, and 14), overlaying each individually recovered footprint against its 
ground truth counterpart. The strong spatial correspondence across all recovered footprints confirms that our framework consistently performs with high fidelity regardless of crowd size or interaction topology.

\subsubsection{Zonal Analysis} Beyond individual footprint recovery, our framework enables downstream analysis of spatial usage patterns over time. As an example, we perform zonal analysis in scenarios 
C2 and C4 to characterize how individuals dwell in different parts of the space and form temporary interaction clusters. Experiment 9 illustrates this capability by capturing two transient cliques formed by small groups pausing to interact. The clique durations predicted by our pipeline are (1.63 s, 1.69 s), closely matching the true durations of (1.38 s, 1.22 s). Similarly, Experiment 11 in Fig.~\ref{fig:all_results_table} captures poster-session dynamics, where pedestrians briefly slow down while scanning displays. The predicted dwell times near the points of interest are (0.36 s, 0.71 s, 0.89 s), closely aligning with the ground-truth values of (0.33 s, 0.50 s, 0.82 s).
\section{Discussion and Future Work}
\subsubsection{Computational complexity} Across all experiments, our framework incurs an average computational overhead of 330\,ms per second of radar data, demonstrating its suitability for online operation. The hypothesis tree has a worst-case complexity of $\mathcal{O}(P^{t_p})$. In practice, this growth is effectively controlled through pruning strategies that discard low-scoring hypotheses, as implemented in our framework. Exploring more advanced pruning or more compact hypothesis representations is an interesting future work direction. 

\subsubsection{Ablation Study} 
\begin{figure}
    \centering
    \includegraphics[width=\linewidth]{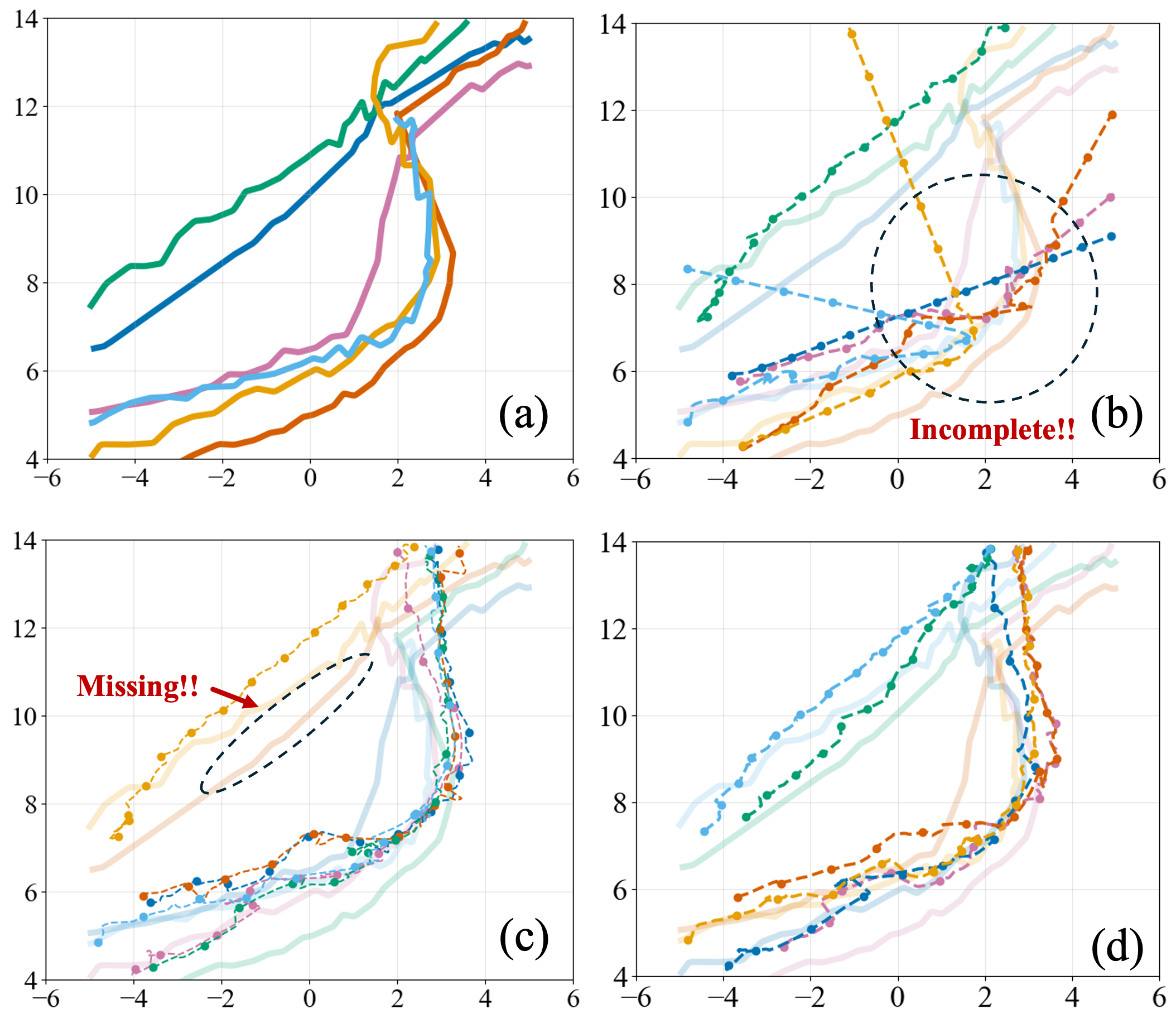}
    \vspace{-18pt}
    \caption{ {\todo{Qualitative} Ablation Study for Experiment 11 (C4, 6 people, Area 4). (a) Ground truth shows the entangled crowd footprints. (b) When all modules are disabled, our framework reduces to a classical state-space estimator, leading to incomplete recovery of footprints during sustained loss of observability. (c) Incorporating only macroscopic crowd semantics, but excluding \textbf{ the physics-aware reasoning module}, captures some general motion but incurs loss of entire footprint paths. (d) Our full framework utilizes both macroscopic semantics and physics-aware reasoning to accurately recover crowd footprints. See color PDF for best viewing.}}
    \label{fig:ablation}
\end{figure}

Our framework is built upon three key components: learned crowd semantics, a principled model of radar observation loss caused by occlusion and merging, and a multi-hypothesis reasoning procedure that systematically integrates these elements with the radar measurements. To assess their individual contributions, we conduct an ablation study as shown in Fig.~\ref{fig:ablation}. Disabling all modules (Fig.~\ref{fig:ablation} (b)) reverts the system to a classical state-space estimation framework. However, this baseline fails to maintain footprint continuity during extended periods of observability loss. Enabling only macroscopic semantics (Fig.~\ref{fig:ablation} (c)) captures the general shape of the crowd footprints but remains inaccurate due to entire motion paths being missed. In contrast, our full pipeline achieves precise footprint recovery and high accuracy (Fig.~\ref{fig:ablation} (d)), while the baseline and intermediate configurations struggle with incompleteness in dense crowds. 

\begin{figure}
    \centering
    \includegraphics[width=\linewidth]{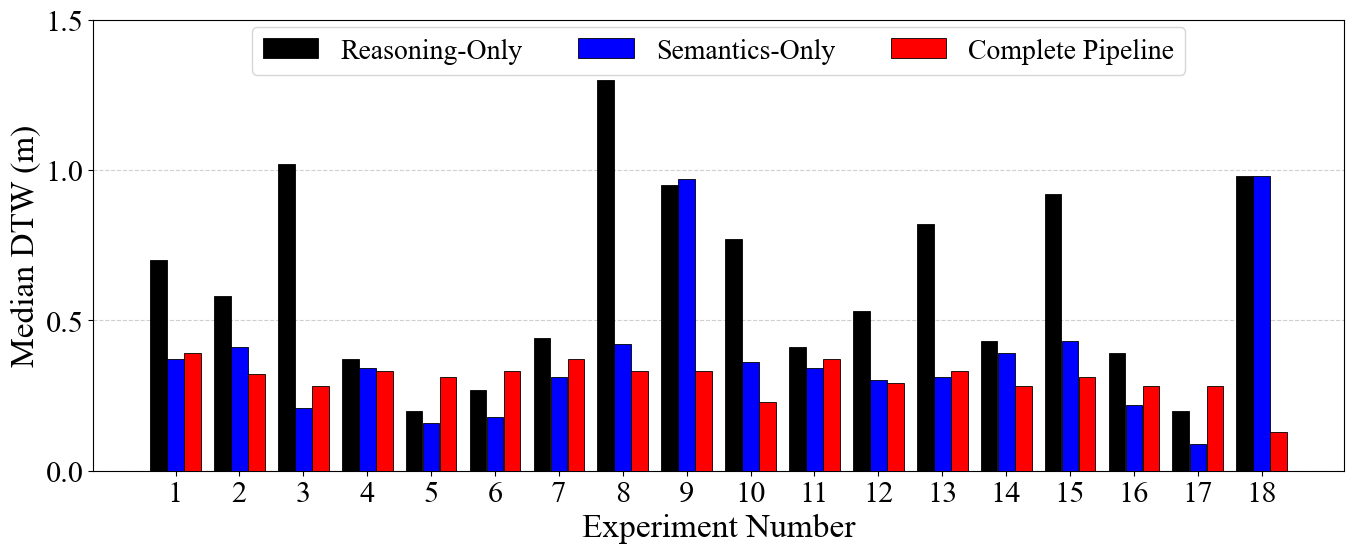}
    \vspace{-22pt}
    \caption{ \todo{Quantitative ablation analysis across all experiments, demonstrating the individual contributions of the physics-aware reasoning module (black) and macroscopic spatial semantics (blue) relative to the complete pipeline (red). See color PDF for best viewing.}}
    \label{fig:quant_ablation}
    \vspace{-22pt}
\end{figure}

\todo{The quantitative results in Fig.~\ref{fig:quant_ablation} further showcase the distinct contributions of each component, with all configurations evaluated under identical experimental conditions. The Semantics-Only pipeline isolates the contribution of macroscopic spatial semantics, exhibiting behavior comparable to the configuration in Fig.~\ref{fig:ablation} (c). The footprints recovered by this module largely follow macroscopic semantic priors, resulting in unrealistic clustering around semantically favorable regions and failing to capture individual pedestrian behavior. 
Furthermore, this module fails to consistently maintain the true crowd size, missing at least one individual in nine of the 15 experiments in which crowd size was not assumed.} \todo{ Consequently, these limitations render the Semantics-Only configuration unsuitable for analyzing crowd behavior. To isolate the complementary contribution of merge and occlusion reasoning, the Reasoning-Only pipeline relies exclusively on these reasoning mechanisms to resolve spatio-temporal entanglement, but does not consider the macroscopic spatial guidance.
Thus, although it maintains the true crowd size for some experiments, the absence of macroscopic spatial guidance can result in substantial spatial inaccuracies in the recovered footprints. Overall, this ablation study demonstrates that both components, as well as their proper integration, are necessary for recovering accurate and physically plausible individual footprints suitable for analyzing spatial crowd behavior.}

\begin{figure}
    \centering
    \includegraphics[width=1\linewidth]{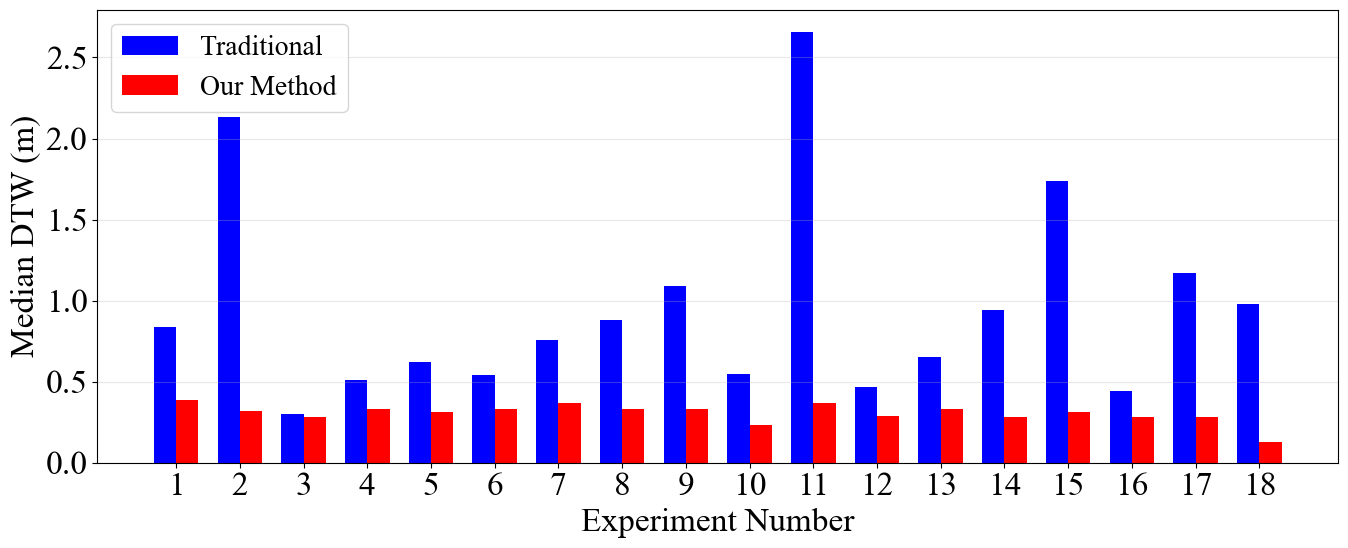}
    \vspace{-22pt}
    \caption{State-Of-the-Art comparison: Quantitative DTW error analysis for both typical MHT trackers and the proposed foundation. Our method significantly outperforming existing approaches, achieving a median DTW error of 0.31 m vs. 0.96 m for the State-Of-the-Art.}
    \label{fig:dtw_comparison}
    \vspace{-20pt}
\end{figure}
\subsubsection{Current Limitations and Motivations for Future Work} \todo{Our framework leverages spatial crowd semantics and principled observability loss reasoning to recover individual footprints under severe merging and occlusion. Fig.~\ref{fig:all_results_table} quantifies this challenge: the average Merge Factor is 29\%, reflecting substantial observability loss and frequent merging across our experiments.  Nevertheless, exceptionally severe and sustained information loss can exceed the proposed foundation's ability to resolve individual footprints. We next discuss a few such examples to motivate future directions. In Experiment 8 (Fig.~\ref{fig:all_results_table}), for instance, two individuals were merged and/or occluded for 73.75\% of the experiment duration, compromising their footprint recovery and yielding DTW errors of 0.47 m and 0.76 m, respectively. These errors are also visually evident in the deviations of their reconstructed footprints from the ground truth in Fig.~\ref{fig:all_results_table}.

A second limitation concerns initialization when individuals enter as a tightly coupled clique. Across our 18 experiments, we provided the total crowd count only for these three cases: Experiments 9, 14, and 15, to initialize individuals uniformly across the entry region. Without this count information, their median DTW errors are 0.77 m, 0.92 m, and 0.39 m, respectively, compared with 0.32 m, 0.28 m, and 0.31 m when the counts are provided. These cases motivate improved initialization strategies or integration with existing crowd-counting methods.}


\subsubsection{State-Of-The-Art Comparison}
In Figs.~\ref{fig:ablation} and \ref{fig:quant_ablation}, we present an ablation study in which each intermediate variant is designed to closely mirror a corresponding class of state-of-the-art approaches. For completeness, we further compare our framework against a traditional Multi-Hypothesis Tracker (MHT)~\cite{blackman1999design} under identical experimental conditions. Although MHT maintains multiple tracking hypotheses, it fundamentally relies on a one-to-one correspondence between observations and tracked individuals. Furthermore, it lacks learned crowd semantics, explicit observability-loss modeling, and principled reasoning over many-to-one and one-to-none observation mappings, all of which are essential for robust operation in dense crowds. As shown in Fig.~\ref{fig:dtw_comparison}, the proposed framework delivers substantial performance gains over MHT across all experimental scenarios. {Specifically, MHT achieves an average median DTW error of 0.96~m, compared with only 0.31~m for the proposed framework, representing more than a threefold increase in error.} These results underscore the inability of conventional methods to cope with the frequent and prolonged spatio-temporal entanglement induced by merging and occlusion in dense crowd environments. 

\subsubsection{Multi-radar implementations} Our framework handles merging, occlusion, and temporary loss of observability well. In scenarios involving very prolonged loss of observability and large numbers of simultaneously unobserved individuals, incorporating multiple radar viewpoints could further enhance robustness. Exploring how to effectively combine information across such perspectives is an interesting direction for future work. Such deployments could also enable behavior analysis over larger spatial areas.


\subsubsection{Person identification}Our framework recovers individual spatial footprints without relying on person identification. Micro-Doppler signatures have demonstrated promise for person identification~\cite{pegoraro2020multiperson}.  However, their discriminability degrades considerably in dense crowds, as reported in the literature~\cite{wang2025grouped}. By recovering disentangled spatial footprints, our framework could provide a natural basis for selectively leveraging micro-Doppler cues over localized regions in future work, enabling identity association in dense crowds while also improving footprint recovery.
\subsubsection{Adaptive hypothesis management} Our framework currently employs score-based pruning to control hypothesis growth. Future work could explore adaptive or learned pruning that dynamically allocate hypotheses to the most ambiguous regions, improving scalability in dense crowds.

\section{Conclusion}
\label{sec:conclusion}
This paper addresses the challenge of recovering individual spatial footprints from spatio-temporally entangled mmWave radar point clouds. We introduced a macro-to-micro framework that bridges learned crowd semantics and physics-informed radar observability modeling, forming the foundation for a semantic-guided multi-hypothesis reasoning procedure that systematically untangles ambiguous radar observations. Evaluated across 18 real-world experiments with crowds of up to (and including) 10, and spanning six crowd topologies in four environments, our framework achieves strong structural consistency with
the ground-truth, with recovered footprints further enabling zonal space usage analysis. Overall, the results demonstrate the viability of mmWave radar as a sensing modality for fine-grained crowd analytics and individual behavior understanding in dense, interaction-rich environments.

\bibliographystyle{ieeetr}
\bibliography{main}

\end{document}